\documentclass[aps,prl,reprint,groupedaddress, showpacs]{revtex4-1}

\usepackage{graphicx}
\usepackage{epstopdf}
\usepackage{amssymb}
\usepackage{amsmath}
\usepackage{bm}
\usepackage{braket}

\newcommand{\insitu}{$in$ $situ$ }

\begin{document}

\title{Coherent inflationary dynamics for Bose-Einstein condensates crossing a quantum critical point} 

\author{Lei Feng, Logan W. Clark, Anita Gaj, Cheng Chin}
\affiliation{James Franck Institute, Enrico Fermi Institute and Department of Physics, University of Chicago, Chicago, IL 60637, USA}

\date{\today}

\begin{abstract}
Quantum phase transitions, transitions between many-body ground states, are of extensive interest in research ranging from condensed matter physics to cosmology~\cite{Subir2011, Morikawa1995,Kibble1980,Vojta2013}. Key features of the phase transitions include a stage with rapidly growing new order, called inflation in cosmology~\cite{Guth2007}, followed by the formation of topological defects~\cite{Kibble1976, Zurek1985, Zurek2014}. How inflation is initiated and evolves into topological defects remains a hot debate topic. Ultracold atomic gas offers a pristine and tunable platform to investigate quantum critical dynamics~\cite{Sadler2006,Anatoli2011,Bloch2008,Dziarmaga2010,Baumann2011,Barnett2011,Lamporesi2013,Nicklas2015,Navon2015,Klinder2015,Meldgin2016,Anquez2016,Clark2016}. We report the observation of coherent inflationary dynamics across a quantum critical point in driven Bose-Einstein condensates. The inflation manifests in the exponential growth of density waves and populations in well-resolved momentum states. After the inflation stage, extended coherent dynamics is evident in both real and momentum space. We present an intuitive description of the quantum critical dynamics in our system and demonstrate the essential role of phase fluctuations in the formation of topological defects. 
\end{abstract}

\maketitle

\begin{figure}
   \centering
    \includegraphics[width=83mm]{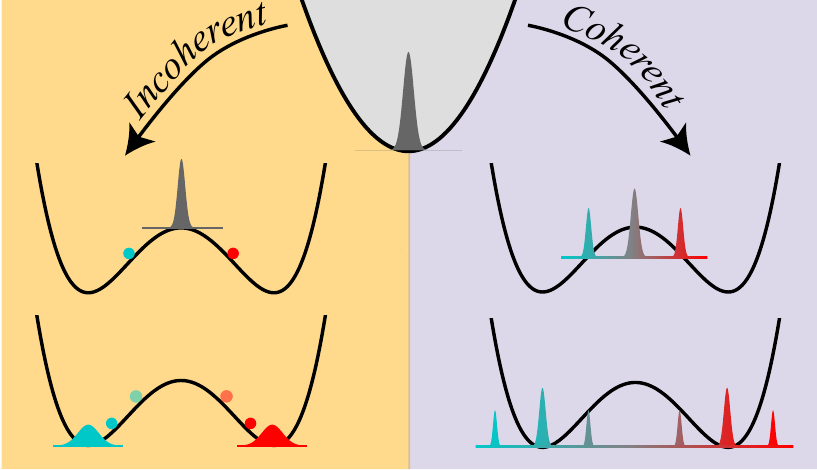}
    \caption{\textbf{Paradigms of dynamics crossing a ferromagnetic quantum critical point } Two scenarios describing the quantum phase transition: (Left) In the incoherent picture, the system broken into locally coherent segments by fluctuations. Each segment evolves independently toward a new ground state. Particles eventually rethermalize at the energy minima to form domains. (Right) In the coherent scenario, the system evolves toward the new ground states with macroscopic coherence extending beyond the domain size. \label{fig_main_1}}
\end{figure}

During a quantum phase transition, a many-body system, originally prepared in the ground state with macroscopic coherence, is suddenly transferred to a metastable state after passing the critical point~\cite{Subir2011,Kibble1976,Dziarmaga2010}.  An example shown in Fig.~\ref{fig_main_1} is a ferromagnetic transition where the $Z_2$ inversion symmetry is broken. How does the system evolve toward the new ground states generally with a different symmetry? One can hypothesize two possible scenarios: 1. Fluctuations break the system into locally coherent segments which evolve toward the new ground states independently. After relaxation, the system form domains with local coherence~\cite{Morikawa1995,Dziarmaga2010}. 2.  Maintaining the macroscopic coherence, the system undergoes a coherent population transfer of particles toward lower energy states. Here fluctuations determine the domain structure but do not destroy the macroscopic coherence.While both scenarios support rapid evolution toward new ground states, the key differences are the time and length scales of the coherence in the dynamical process.


In this paper, we report the observation of coherent inflationary dynamics in an atomic Bose condensate driven across a quantum critical point. Our experiment is based on cesium Bose-Einstein condensates loaded into a one-dimensional phase-modulated optical lattice \cite{Harry2013}. The modulation translates the lattice periodically with displacement $\Delta x = \frac{s}{2}\sin\omega t$, where $s$ is the shaking amplitude and $\omega$ is the shaking frequency. Shaking hybridizes the ground and excited Bloch bands and results in an effective dispersion $\varepsilon_q$ for the condensate \cite{Harry2013}, where the lowest energy state at quasi-momentum $q$ = 0 bifurcates into two ground states at $+q^*$ and $-q^*$ (named pseudo-spin up and down), when $s$ exceeds a critical value $s_c$. When the system is driven across the critical point in finite time, domains of pseudo-spins form in accordance with universal Kibble-Zurek scaling \cite{Clark2016} and excitations within a domain display a roton dispersion~\cite{Harry2015}, however, a complete understanding of the processes that underlie the quantum critical dynamics remains evasive.

\begin{figure*}
\centering
\includegraphics[width=180mm]{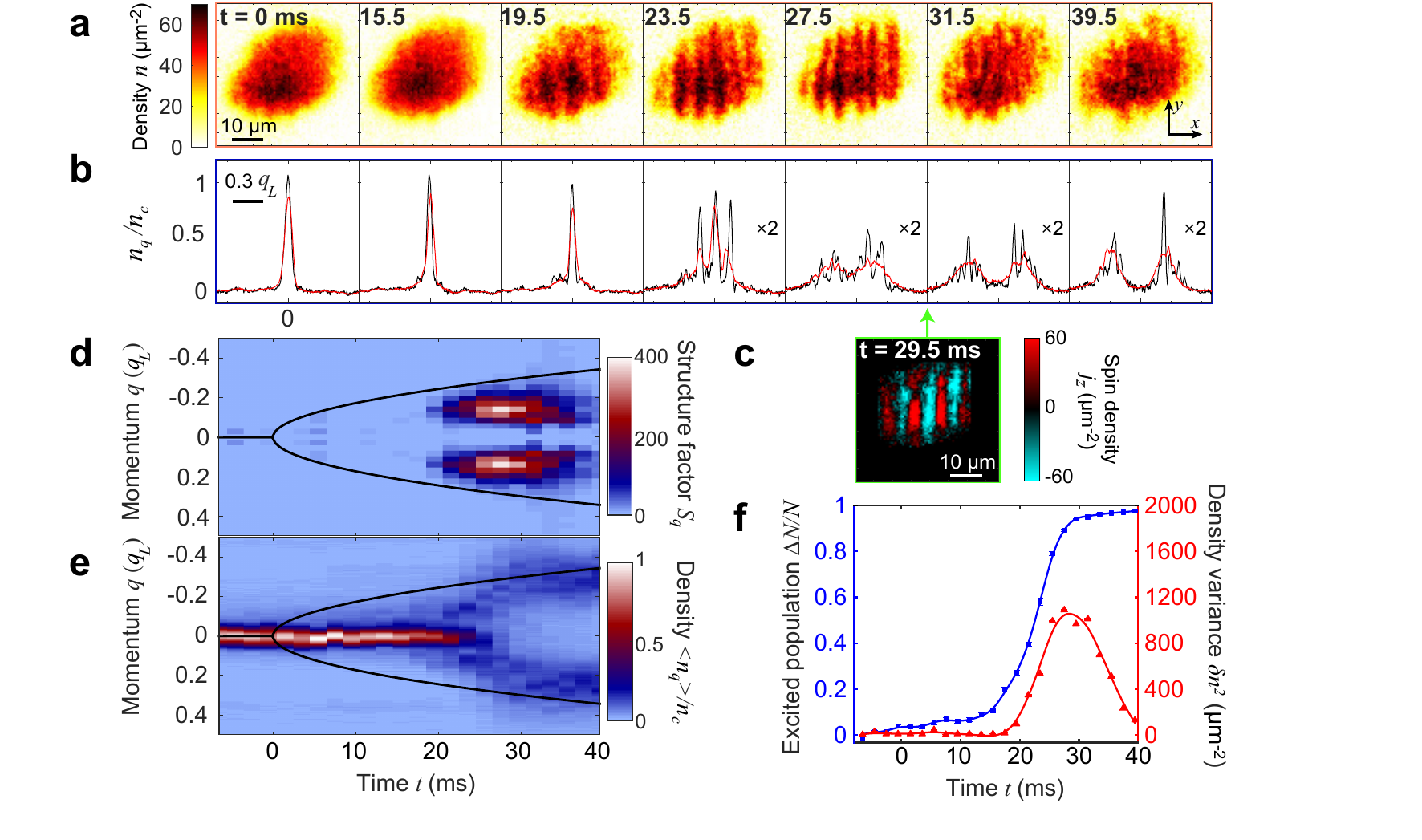}
\caption{\textbf{Development of density waves and momentum space population across the quantum critical point } We linearly ramp up the shaking amplitude $s$ with a ramp rate $\dot{s}$  = 0.64 nm/ms across the critical point at time $t$~=~0. \textbf{a,} Single shot \insitu images of the condensate. \textbf{b,} Momentum distribution $n_q$ from time-of-flight measurement(black). Here $n_c$ is the averaged peak density in the momentum space of unshaken condensates. Averaging over repeated experiments gives two broad peaks centered around $q$ = 0 (red). \textbf{c,} Domain structure from reconstruction \cite{Clark2016}, where $j_z = n_{q^*} - n_{-q^*}$ is the spin density. \textbf{d,} The density structure factor $S_q~=~\left<\Delta n_q^2\right>/N$, extracted from the Fourier transform of the density fluctuation $\Delta n(x)= n(x) - \left<n(x)\right>$ integrated along $y$ axis. Here $N$ is the total atom number and $\langle .\rangle$ indicates an average over repeated measurements. Peaks appear at $\pm q_d$~=~$\pm$0.14~$q_L$ with $q_L = \pi\hbar/d$ being the lattice momentum and $d$ being the lattice period. \textbf{e,} The averaged population distribution $\langle n_q\rangle$ in momentum space. Solid black curves in both \textbf{d} and \textbf{e} show the instantaneous, theoretical ground state momenta $\pm q^*$. \textbf{f,} Fractional population excited out of $q=$ 0 state (blue square) and the density variance $\delta n^2$ from integrating the structure factor $S_q$. Solid lines are guides to the eye.
	\label{fig_main_2}}
\end{figure*}

To reveal the nature of the quantum phase transition, we exploit three schemes to analyze the critical dynamics of the condensate: 1. \insitu imaging to record the atomic density profile, 2. time-of-flight with focusing technique~\cite{Shvarchuck2002} to probe the momentum space distribution $n_q$, and  3. pseudo-spin reconstruction to reveal domain structure~\cite{Clark2016}. An example is shown in Fig.~\ref{fig_main_2}. Here we linearly ramp up the shaking amplitude and interrupt the ramp at time $t$ after passing the critical point to probe the system with the 3 methods.

We observe two key features indicating coherent evolution. First, from \insitu images, density wave emerges about 20~ms after passing the critical point. Quantified with the density structure factor $S_q$ \cite{Chenlung2011}, the density wave shows an almost fixed wavenumber. Second, from time-of-flight images, atomic population forms sharp side peaks in individual sample; over repeated measurements the side peaks average to broader features. These observations suggest that atoms occupy a coherent superposition of well-defined momentum states and the density wave emerges from their interference. Though the density wave diminishes after 30~ms, the persistent narrow momentum peaks in atomic population $n_q$ suggest a long-lasting coherence. In addition, the period of the density waves approximately matches twice the averaged domain size. Both features will be further discussed in later paragraphs.

A more comprehensive analysis of the density wave and the population distribution in momentum space suggests that the system evolution can be separated into two stages: inflation and relaxation. To see this, we evaluate the density variance $\delta n^2 = \int dq S_q$ from \insitu images as well as  total population in finite momentum states  $\Delta N = \sum_{q>0}N_q$ from time-of-flight measurements, where $N_q$ is the total atom number in $\pm q$ states. For short times after the phase transition, both quantities show a characteristic exponential-like growth; we name this period the inflation stage, see Fig.~\ref{fig_main_2}\textbf{f}. After the inflation, all atoms relax toward non-zero momentum states while the density wave diminishes. In the following, we investigate the two stages separately.

The exponential growth of excitations can be theoretically understood based on dynamical instability of the condensate~\cite{Morsch2006}. Shortly after passing the quantum critical point, the $q=0$ state remains macroscopicly occupied, which justifies the Bogliubov approximation. Because of the inverted dispersion, the many-body Hamiltonian cannot be diagonalized with bosonic field operators~\cite{Anglin2003}. Instead we can write the Hamiltonian as (Supplementary Information)
\begin{equation} \label{eq:ground_band}
H = \sum_{q>0}\hbar\lambda_q(\hat{\iota}^\dagger_{q}\hat{\iota}^\dagger_{-q}+\hat{\iota}_{q}\hat{\iota}_{-q}),
\end{equation}
\noindent where $\hat{\iota}^\dagger_q$ and $\hat{\iota}_q$ are the bosonic creation and annihilation operators of an inflaton with momentum $q$ and growth rate $\lambda_q = \sqrt{-\varepsilon_q(2\mu+\varepsilon_q)}/\hbar$, $\mu$ is the chemical potential and $\hbar$ is the reduced Planck constant. It is important to emphasize that only modes with negative kinetic energy $\varepsilon_q<0$ acquire the inflationary dynamics. Based on the Hamiltonian, the excited populations increase exponentially in the inflation phase according to (Supplementary Information)
\begin{equation}
N_q(t)+1 = [N_q(0)+1]\cosh2\lambda_qt,\label{observable}
\end{equation}
and the structure factor $S_q(t)=(-\varepsilon_q/\mu)[N_q(t)+1]$ near the critical point. This result explains the similar exponential-like growth of both observables in Fig.~\ref{fig_main_2}\textbf{f}. 

To further test the inflation theory, we perform quench experiments by suddenly driving the system across the critical point, and measure the growth rate of the population in different momentum modes. Right before the quench, we seed a small initial population in the desired momentum states $\pm q'$ by imprinting a sinusoidal phase pattern on the condensate $\delta\phi\sin(q'x/\hbar)$. Here $\delta\phi$ is the seed amplitude and the wavenumber $q'/\hbar$ is externally controlled (See Methods). 

\begin{figure}
	\centering
	\includegraphics[width=83mm]{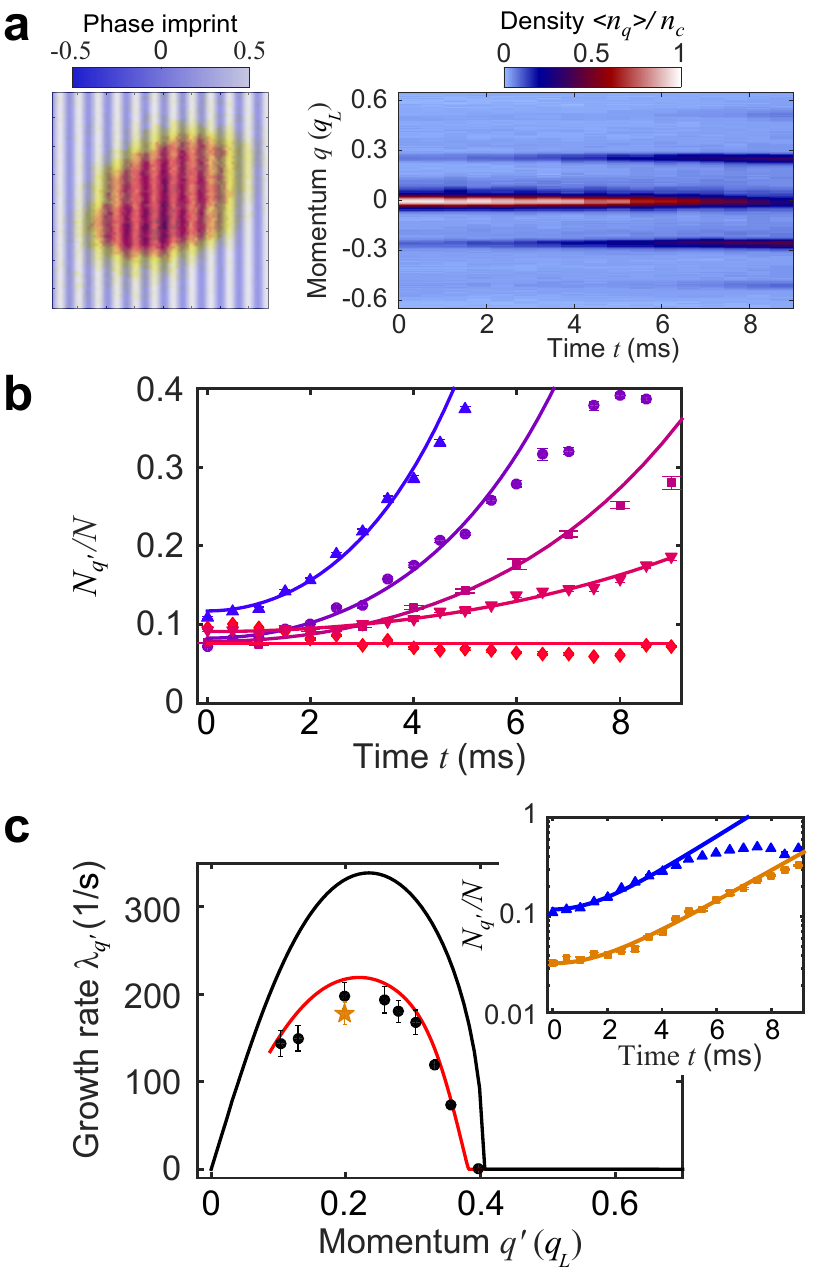}
	\caption{\textbf{ Growth of excitations during the inflation phase} Condensates are quenched from shaking amplitude $s_c=$~13.1~nm to $s=$~25~nm, where the new ground states are at $\pm q^*$ = $\pm$0.24~$q_L$. \textbf{a,} At $t=0$, we quickly imprint a phase modulation in 20~$\mu$s on the condensate with a seeding momentum $q'$ = 0.26~$q_L$ (left). Subsequent time of flight measurements reveal two side peaks emerging at $\pm q'$ (right). \textbf{b,} the fractional population in both side peaks $N_{q'}/N$ evolves for different seeding momentum: $q'$ = 0.19 (triangle), 0.30 (circle), 0.33 (square), 0.36 (inverted triangle) and 0.40~$q_L$ (diamond), from blue to red. Solid lines are fits using Eq.~(\ref{observable}) to extract the growth rate $\lambda_{q'}$. \textbf{c,} The growth rates for seeded (black) and unseeded experiments (orange star) are compared with Bogoliubov theory (black line)and numerical simulation (red line). Inset compares the growth for seeded experiment with $q'$ = 0.19~$q_L$ (blue) and the unseeded quench experiment (orange). \label{fig_main_3}}
\end{figure}

After seeding, the condensate quickly grows two side peaks exactly at the seeding momentum $\pm q'$ (Fig.~\ref{fig_main_3}). To extract the growth rate, we monitor the population in the momentum states. The population grows exponentially in the beginning but reaches a maximum at a later time when the population in the $k=0$ state is depleted. We fit the fast growing interval right after the quench according to Eq.~(\ref{observable}) and compare the growth rate to the prediction. Our measured growth rates qualitatively agree with the Bogoliubov result. We find quantitative agreement with our numerical simulation based on Gross-Piteavskii equation which incorporates the depletion of the condensate (Supplementary Information). In particular, we confirm that only modes $|q|\leq$ 0.4~$q_L$ with kinetic energy $\varepsilon_{q}<0$ exponentially grow and the fastest growth appears at momentum $q\approx\pm q^*$.

Remarkably, in the absence of seeding, the sample spontaneously grows momentum peaks near $\pm q^*$ with a growth rate very close to that seeded at a similar momentum (Fig.~\ref{fig_main_3}\textbf{c}). For unseeded samples, many momentum modes can in principle be populated by quantum or thermal fluctuations and then amplified by inflation. The dominance of the modes near $\pm q^*$ can be understood since they have the highest growth rate and become dominant during inflation.

\begin{figure}
	\centering
	\includegraphics[width=83mm]{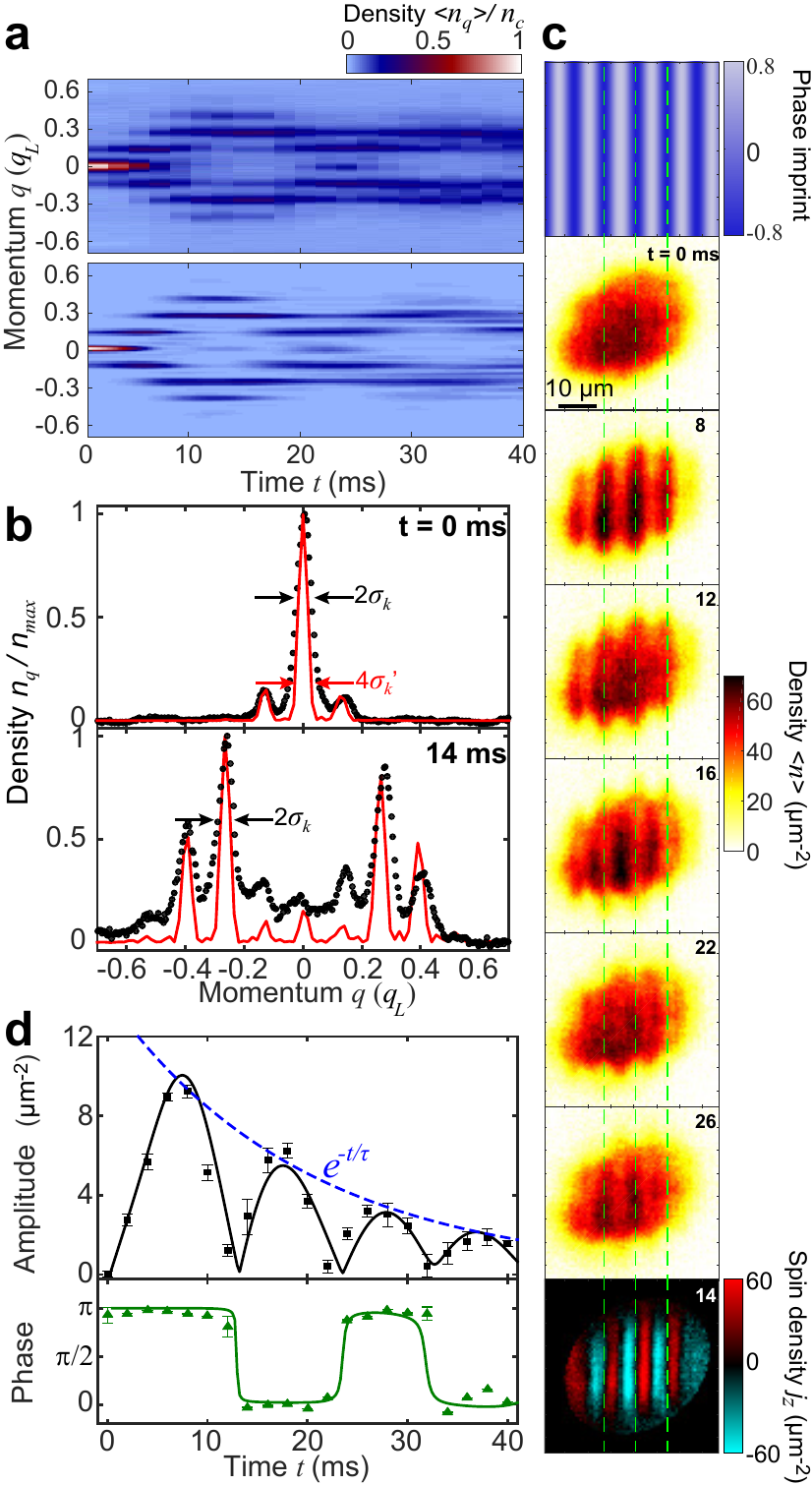}
	\caption{\textbf{Coherent quantum critical dynamics} The condensates are seeded at wavenumber $q'$ = 0.13~$q_L$ and quenched from shaking amplitude $s_c=$~13.1~nm to $s=$~25~$nm$. \textbf{a,} Coherent oscillations in momentum space (top: experiment, bottom: numerical calculation). \textbf{b,} Line cuts of the experimental data at $t$ = 0 and 14~ms (black dot). The solid red lines are from numerical calculations. The experimental peaks in both cuts show similar root-mean-square diameters of $\sigma_{q}$ =~0.026~$q_L$ and 0.028~$q_L$ respectively from Gaussian fits. The numerical calculation shows $\sigma'_{q}$~= 0.015~$q_L$, determined by the sample size. \textbf{c,} Oscillation of the density wave and the domain structure at $t=14$~$ms$. Both density waves and domains appear aligned with the seed pattern (green dashed line). \textbf{d,} Amplitude (black square) and phase (green triangle) of the density wave are compared with the numerical calculation (solid lines). A settling time $\tau$~=~20~ms is extracted from the decay of the envelope function (blue dashed line).
    \label{fig_main_4}}
\end{figure}

Following the inflation stage, the condensates display persistent coherent dynamics in both time-of-flight and \insitu measurements (Fig.~\ref{fig_main_4}). After the rapid growth of the population at seeded momentum $\pm q'$, the system generates higher order harmonics at $\pm 2q', \pm 3q'...$, and the atomic populations are coherently transferred between these momentum states. The emergence of higher harmonics is due to nonlinear mixing of the matter waves and can be well described based on our numerical model. An example at $t$ = 14~ms shows multiple side peaks that conform to the simulation. Intriguingly, the individual momentum peaks are as narrow as the zero momentum peak of the original condensate; the widths are only limited by the detection resolution. The narrow momentum peak indicates long coherence length based on uncertainty principle. By comparing our measurement to the simulation, we conclude the lower bound of the coherence length to be 15~$\mu$m, which is much greater than the averaged domain size of 4.1(1)~$\mu$m (Supplementary Information).


Together with the dynamics in the momentum space, density waves in seeded samples also display coherent oscillations in quench experiments (Fig.~\ref{fig_main_4}\textbf{c}). The density wave appears aligned to the imprinted pattern, and its phase displays multiple alternations (Fig.~\ref{fig_main_4}\textbf{d}) which are synchronized with oscillations of the population in momentum space. The contrast of the density wave slowly decays with a time constant of $\tau$~=~20~ms. Both the alternation and the decay are in good agreement with our simulation. Finally, we find that domains are fully formed as early as $t=14$~ms, and remain constant afterward. Importantly, the domain structure is deterministic in the seeded experiments, and the domain walls line up with the density wave yielding a domain size half the period of the density wave.

The observed coherent dynamics can be understood based on a simple physical picture. Phase imprinting across the condensate locally breaks the inversion symmetry by providing a local momentum kick $q(x) = \partial_x\phi(x)$, where $\phi(x)$ is the phase of the condensate wave function. Within one period of the imprinted phase pattern, the sign of the local momentum flips twice, resulting in two neighboring domains with opposite momenta. After the momentum kick, atoms in neighboring domains can flow toward or away from each other determined by the group velocity $v_g(x) = d\varepsilon_q/dq$, leading to the observed density peaks and troughs. Since density waves cost energy in a BEC with repulsive interactions, the atom flow reverses after half an oscillation period, yielding the phase alternation of the density wave. 

The decisive role of phase imprinting in the real and momentum space dynamics and domain structure indicates the importance of phase fluctuations in quantum critical dynamics. Together with the emergence of density wave and atomic occupation in well-resolved momentum states, we present strong evidence supporting the coherent scenario of the quantum phase transition in our system. Furthermore, the phase imprinting technique can find new applications to engineering desired structure of domain walls, which will enable future study on the dynamics and interactions of topological defects.

\begin{acknowledgments}
We thank E. Berg, Q. Zhou and B. M. Anderson for helpful discussions. L. W. C. was supported by Grainger fellowship. A. G. is supported by Kadanoff-Rice fellowship. This work was supported by NSF Materials Research Science and Engineering Centers (DMR-1420709), NSF grant PHY-1511696, and Army Research Office-Multidisciplinary Research Initiative grant W911NF-14-1-0003. The data presented in this paper are available upon request to C.C. (cchin@uchicago.edu). 
\end{acknowledgments}

\noindent\\

\noindent\textbf{Methods}

\noindent\textbf{Lattice loading.} We utilize three-dimensional Bose-Einstein condensates of 30,000 cesium atoms confined in an optical dipole trap. The trap is tightly confined in the gravity direction with a trapping frequency of $2\pi\times$226~Hz. Trapping frequencies in the two in-plane directions are $2\pi\times$6 and $2\pi\times$9~Hz. The $s$-wave scattering length is 2.6~nm. We adiabatically load the BEC into a one-dimensional optical lattice with a depth of 8.9~$E_R$ and period $d$~=~532~nm, where $E_R$ = $h\times$ 1.3~kHz is the recoil energy.

\noindent\textbf{Shaken lattice.} We periodically translate the lattice by sinusoidally modulating the phase of one of the lattice beams. The shaking frequency is fixed to $\omega$~=~$2\pi \times$8 kHz, which is 2$\pi\times$0.87~kHz above the gap at zero-momentum between the ground and the first excited Bloch band in the lattice. Given the lattice depth and the shaking frequency, the critical shaking amplitude is $s_c$~=~13.1~nm. 

\noindent\textbf{Phase imprinting.} We imprint the phase pattern across the condensate using a digital micromirror device (DMD) with a 795 nm laser. To ensure a sinusoidal modulation, we set a grating pattern on the DMD with twice of the desired period $2\pi\hbar/q'$ and only let the $\pm$1 orders from the diffraction pass in the Fourier plane. The diffracted light interfere on the atom giving a clean sinusoidally varying potential. The imprinting pulse lasts for 20 to 40~$\mu s$, which is very short compared to the condensate and lattice time scale.

\section{Supplementary Information}
\subsection{Experiment technique and data analysis}

\noindent\underline{\textbf{Time-of-flight imaging with focusing}}

In order to map out the quasi-momentum states of the condensates with high resolution, we perform the time-of-flight (TOF) measurement with focusing technique\cite{Shvarchuck2002}.  Firstly, to avoid atom collisions, we reduce the scattering length to zero by switching the magnetic field right before time-of-flight. At the same time, we increase the magnetic field gradient to properly levitated the atoms into our imaging plane while we turn off the dipole traps. During time-of-flight, all the dipole traps are turned off except the one along lattice direction. By carefully tuning the intensity of the dipole beam, we are able to focus the atoms in the same momentum state into a sharp peak after a quarter of the trapping period of 60~ms. With this technique, we are able to reach a resolution in the momentum space to $\delta q$ = 0.08~$q_L$, only limited by the anharmonicity of our trap.\\

\noindent\underline{\textbf{Momentum state population and growth rate}}

In the ramp experiment, we want to extract the excited population $\Delta N$ in the momentum states $q\neq$~0. Firstly, we perform the focusing time-of-flight measurements on pure Bose-Einstein condensates and characterize the atom distribution in momentum space $n_0(x)$. In the following moments during the ramp, we fit the center interval, $[-\delta q,\delta q]$, of the atom distribution $n_q$ using the characterized profile to quantify the atom population $N_0$ in $q$~=~0 state. As a result, the excited population $\Delta N = N - N_0$ with $N$ the total number of atoms.

In the quench experiments, to extract the atom population $N_{q'}$ in the momentum states $\pm q'$, we integrate the populations in the momentum states within the intervals $[q'-\delta q, q'+\delta q]$ and $[-q'-\delta q, -q'+\delta q]$. To extract the inflaton growth rate, we fit the fast growing interval of our data in early times where, where $N_{q'}/N\leq$ 0.3, to $A\cosh(2\lambda_{q'}t)$ and both $A$ and $\lambda_{q'}$ are simultaneously determined.\\

\noindent\underline{\textbf{Amplitude and phase of the density wave} } 

To accurately extract the amplitude and phase of the density wave, we have to consider the density inhomogeneity in the harmonic trap. This density inhomogenity affects little the amplitude but significantly influences the phase due to nonuniform chemical potential. To minimize this effect, we choose a rectangular subset of the center part of the atom cloud, where local density is no less than 80\% of the peak density of the condensate. We integrate over the non-lattice direction and obtain density distribution along the lattice direction $\langle n(x)\rangle$, averaged over multiple shots. The density modulation $\langle\Delta n(x)\rangle$ is obtained by subtracting $\langle n(x)\rangle$ from the initial density distribution $\langle n_{0}(x)\rangle$ right before the shaking just starts. The amplitude and phase are then given by the Fourier transform of $\langle\Delta n(x) \rangle$.\\

\noindent\underline{\textbf{Coherence length estimation} } 

Here we extract a lower bound of the coherence length based on uncertainty principle in Fig.~4~B. Given the mean square width of the atom population peaks in momentum space $\sigma_k$, the coherence length of the system is then $l_\phi \propto h/\sigma_k$. In our simulation, based on fully coherent condensates with coherence length equals to the system diameter of 30~$\mu m$, the mean square width $\sigma_k'$ is almost half of that in the corresponding experiments. This indicates that the coherence length $l_\phi>$~15~$\mu m$ in the experiment.

\subsection{Theory of Inflation}

\noindent\underline{\textbf{Many-body Hamiltonian}}

Here we start with the many-body Hamlitonian in the second quantization form \cite{Pethick2008},
\begin{eqnarray}
H=\sum_q\varepsilon_q \hat{a}_q^{\dagger}\hat{a}_q + \frac{U_0}{2V} \sum_{q,q',p}  \hat{a}_{q+p}^{\dagger} \hat{a}_{q'-p}^{\dagger} \hat{a}_q \hat{a}_{q'},\label{many_body_hamiltonian}
\end{eqnarray}
\noindent where $\varepsilon_q$ is the single atom dispersion, $U_0$ is the two-body interaction energy, $V$ is the volume of the system and $\hat{a}^{\dagger}_{\pm q}$ and $\hat{a}_{\pm q}$ are bosonic creation and annihilation operators of an atom with momentum $\pm q$. Shaking of the optical lattice dramatically change the single particle dispersion and in the ferromagnetic phase it is given by
\begin{eqnarray}
\varepsilon_q=\epsilon(\frac{q^2}{q^*}-1)^2-\epsilon,
\end{eqnarray}
where $\epsilon$ is the kinetic energy barrier hight. Thus the dispersion is in a double-well shape in momentum space and has two minima at $q=\pm q^*$ sparated by the kinetic energy barrier~\cite{Harry2013, Clark2016}.\\

\noindent\underline{\textbf{Hamiltonian of inflaton and solution}}

Under Bogliubov approximaton, we introduce a new quasi-particle field, inflaton, to rewrite the Hamiltonian of Bose-Einstein condensates. At the beginning of the phase transition, we assume that the $q = 0$ state is macroscopically occupied. Thus the condensate consists of $N$ atoms and has a chemical potential $\mu=U_0N/V$. The many-body Hamiltonian Eq.(\ref{many_body_hamiltonian}) reduces to
\begin{equation}
\begin{matrix}
H&=&\frac12 N\mu+\sum_{q>0}(\mu+\varepsilon_q)(\hat{a}_q^{\dagger}\hat{a}_q+\hat{a}_{-q}^{\dagger}\hat{a}_{-q})\\
\\
& &+ \sum_{q>0} \mu(\hat{a}_q^{\dagger}\hat{a}_{-q}^{\dagger}+\hat{a}_{q}\hat{a}_{-q}).
\end{matrix}
\end{equation}
Conventional Bogoliubov transformation to diagonalize the Hamiltonian using bosonic operators fails in our case because of $\varepsilon_q<0$. Instead, we adopt a different approach by introducing the following transformation
\begin{eqnarray}
\hat{\iota}_{\pm q}&=& u_q \hat{a}_{\pm q}+\nu_q \hat{a}_{\mp q}^{\dagger},
\end{eqnarray}
\noindent where $\hat{\iota}^\dagger_{\pm q}$ and $\hat{\iota}_{\pm q}$ are the creation and anihilation operators of an inflaton with momentum $\pm q$ and the coefficients $u_q, \nu_q>0$ for $\varepsilon_q<0$ satisfy
\begin{eqnarray}
u_q^2&=& \frac12 (\frac{\mu}{\hbar\lambda_q}+1)   \\
\nu_q^2&=& \frac12 (\frac{\mu}{\hbar\lambda_q}-1)   \\
\hbar\lambda_q&=&\sqrt{-\varepsilon_q(2\mu+\varepsilon_q)}.
\end{eqnarray}
The inflaton field operators obeys bosonic commutation relations: $[\hat{\iota}_{q},\hat{\iota}_{q'}]=[\hat{\iota}^{\dagger}_{q},\hat{\iota}^{\dagger}_{q'}]=0$ and $[\hat{\iota}_{q},\hat{\iota}^{\dagger}_{q'}]=\delta_{qq'}$.

As a result, the Hamiltonian reduces to
\begin{equation}
\begin{matrix}
H=\sum_{q>0}\hbar\lambda_q(\hat{\iota}_q^{\dagger}\hat{\iota}_{-q}^{\dagger}+\hat{\iota}_q\hat{\iota}_{-q})\\
\\
+\frac12 N\mu-\sum_{q>0}(\mu+\varepsilon_q).
\end{matrix}
\end{equation}
Beside the two constant terms, the Hamiltonian shows that inflatons are created and annihilated in pairs with opposite momentum. Here the inflation dispersion $\lambda_q$ is related to the growth rate of the inflatons. 

To show the exponential growth, we look at the  dynamics of the inflaton in the Heisenberg picture, which yields
\begin{eqnarray}
\partial_t\hat{\iota}_{\pm q}(t)&=&\frac i{\hbar}[H,\hat{\iota}_{\pm q}(t)]=-i\lambda_q\hat{\iota}_{\mp q}^{\dagger}(t)\\
\partial_t\hat{\iota}^{\dagger}_{\pm q}(t)&=&\frac i{\hbar}[H,\hat{\iota}^{\dagger}_{\pm q}(t)]=i\lambda_q\hat{\iota}_{\mp q}(t).
\end{eqnarray}
The solutions of above equation is  
\begin{eqnarray}
\hat{\iota}_{\pm q}(t)&=&\hat{\iota}_{\pm q}(0)\cosh\lambda_q t -\hat{\iota}^{\dagger}(0)_{\mp q}i\sinh\lambda_q t\\
\hat{\iota}^{\dagger}_{\pm q}(t)&=&\hat{\iota}^{\dagger}_{\pm q}(0)\cosh\lambda_q t+\hat{\iota}_{\mp q}(0)i\sinh\lambda_q t.
\end{eqnarray}
Based on these solutions, the inflaton population $m_{\pm q}(t)\equiv<\hat{\iota}_{\pm q}^{\dagger}(t)\hat{\iota}_{\pm q}(t)>$ evolves according to
\begin{equation}
\begin{matrix}
m_{\pm q}(t)=m_{\pm q}(0)\cosh^2\lambda_q t + m_{\mp q}(0)\sinh^2\lambda_q t\\
+\sinh^2\lambda_q t,
\end{matrix}
\end{equation}
where the first two terms on the right-hand-side correspond to Bose stimulation and the last term originates from spontaneous emission of inflatons. 

We further simplify the results by defining the total inflaton population in $\pm q$ modes $M_k = m_q+m_{-q}$ and have
\begin{eqnarray}
M_q(t)+1=[M_q(0)+1]\cosh2\lambda_q t.
\end{eqnarray}
From this result we see that the inflaton population, including the contribution from spontaneous emission, grows exponentially with a rate of $2\lambda_q$.\\

\noindent\underline{\textbf{Experimental observables}}

The exponential growth of inflaton fields is reflected from the momentum population in the time-of-flight measurements, and the structure factor of the density wave. First of all, we define the population in momentum state $\pm q$ as $N_q = \langle\hat{a}^\dagger_{q}\hat{a}_{q}\rangle + \langle\hat{a}^\dagger_{-q}\hat{a}_{-q}\rangle$. To calculate the atom population from inflatons, we engage the inverse inflaton transformation $\hat{a}_{ \pm q}= u_q \hat{\iota}_{\pm q}-\nu_q \hat{\iota}_{\mp k}^{\dagger}$. We thus rewrite the atom population as

\begin{equation}
N_q + 1 = \frac{\mu}{\hbar\lambda_q}\left(M_q+1\right)-\frac{\mu+\varepsilon_q}{\hbar\lambda_q}\left(\langle\hat{\iota}_{q}\hat{\iota}_{-q}+\hat{\iota}^\dagger_{q}\hat{\iota}^\dagger_{-q}\rangle\right).\label{population}
\end{equation}
Since $\hat{\iota}_{q}\hat{\iota}_{-q}+\hat{\iota}^\dagger_{q}\hat{\iota}^\dagger_{-q}$ commutes with the Hamiltonian and does not vary with time,  the number of atoms in an inflaton is given by $\partial N_q/\partial M_q = \mu/\hbar\lambda_q$ .

Secondly, the structure factor, defined as the Fourier transform of the density-density correlation function \cite{Landau2008}, can be expressed in terms of the correlations in the momentum space as
\begin{equation}
S_q=\frac1N \sum_{p,p'}\langle\hat{a}^{\dagger}_{p+q}\hat{a}_p\hat{a}^{\dagger}_{p'-q}\hat{a}_{p'}\rangle.
\end{equation}
For small number of excitations $\langle\hat{a}^\dagger_q\hat{a}_q\rangle\ll \langle\hat{a}^\dagger_0\hat{a}_0\rangle\approx N$ and $\varepsilon_q<0$ , one can rewrite $S_q$ as
\begin{equation}
S_q=\frac{-\varepsilon_q}{\hbar\lambda_q}\left(M_q+1 + \langle \hat{\iota}_{q}\hat{\iota}_{-q}+ \hat{\iota}_{q}^{\dagger}\hat{\iota}^\dagger_{-q} \rangle\right).\label{structure}
\end{equation}

Based on Eqs.(\ref{population}) and (\ref{structure}), we futher determine the time evolution of the observables. In the begining of phase transition, we assume that there is no net source of correlated inflatons in a regular Bose-Einstein condensate
\begin{equation}
\langle\hat{\iota}_{q}\hat{\iota}_{-q}\rangle = \langle\hat{\iota}^\dagger_{q}\hat{\iota}^\dagger_{-q}\rangle = 0.
\end{equation}
As a result, the initial value of inflaton population at $t$ = 0 is
\begin{equation}
M_q(0)+1 = \frac{\hbar\lambda_q}{\mu}[N_q(0)+1].
\end{equation}
Finally, we obtain the time evolution of $N_q(t)$ and $S_q(t)$,
\begin{eqnarray}
N_q(t)+1 &=& [N_q(0)+1]\cosh(2\lambda_qt)\\
S_q(t) &=& \frac{-\varepsilon_q}{\mu}[N_q(0)+1]\cosh(2\lambda_qt).
\end{eqnarray}

\subsection{Numerical Simulation}

Our numerical simulation is based on the Gross-Pitaevskii equation with an effective model in a one-dimensional lattice. In this model, space is discretized into lattice sites and kinetic motion is replaced by hopping between neighboring sites. We define $\psi_j(t)$ the value of the condensate wave function of site $j$ at time $t$ and the evolution is given by 

\begin{equation}
\begin{matrix}
i\gamma\hbar\frac{\partial}{\partial t}\psi_j(t) &=&-\sum_{n=1}^{n_T}t_k[\psi_{j-n}(t)+\psi_{j+n}(t)]\\
\\
&+&\left[\frac{1}{2}m\omega^2(jd)^2 +g|\psi_j(t)|^2-\mu\right]\psi_j(t)
\end{matrix}
\end{equation}

Here $t_n$ is the tunneling energy between two lattice sites spatially separated by $n$ sites, $m$ is the atomic mass of Cs, $\omega$ is the harmonic trapping frequency, $g$ is the coupling constant, $\mu$ is the chemical potential , $d$ is the lattice period and $n_T$ is the truncation number that shall be discussed later. $\gamma$ is a parameter that controls the evolution in either imaginary time ($\gamma$ = $i$) or real time ($\gamma$ = 1). Particularly in imaginary time, evolution of the equation gives the ground state wave function of the system. In our simulation of inflation, we keep a small imaginary part by setting $\gamma = \cos\varphi+i\sin\varphi$ with $\varphi$ = 0.015, which accounts for the dissipation in our system.

Tunneling energy $t_n$ is determined from the Bloch band structure. In our experiment, all the atoms dominantly occupy the ground band even with lattice shaking. The tunneling energy is then the inverse Fourier transform of the ground band dispersion $t_n = \sum_{q}e^{-inq/\hbar}\varepsilon_q$. In a non-shaken lattice, the nearest-neighbor tunneling $t_1$ is the only significant term and sufficient to capture most of the physics. However, in a shaken lattice, where the ground band dispersion is significantly modified, higher order tunneling become significant as well. In our simulation, we found that $n_T$ = 3 is a good truncation number and $t_n$ with $n>3$ are negligible.

We perform the simulation for our seeded quench experiment as the following. We first evolve equation in imaginary time to reach the ground state wave function in a non-shaken lattice. To implement the seed, we multiply this ground state wave function by a spatially-varying phase factor $e^{i\delta\theta(j)}$ with $\delta\theta(j) = \delta\phi \sin(\frac{q'}{\hbar}jd)$ as the initial wave function for the next step, where $\delta\phi$ is the seed amplitude and $q'$ is the seed momentum. To simulate the quench, we now evolve the initial wave function with a new set of tunneling energies in a shaken lattice. Later all the observables are extracted from the wave function in the same way as we did in our data anlaysis. 

\begin{figure}
\raggedright
\includegraphics[width=83mm]{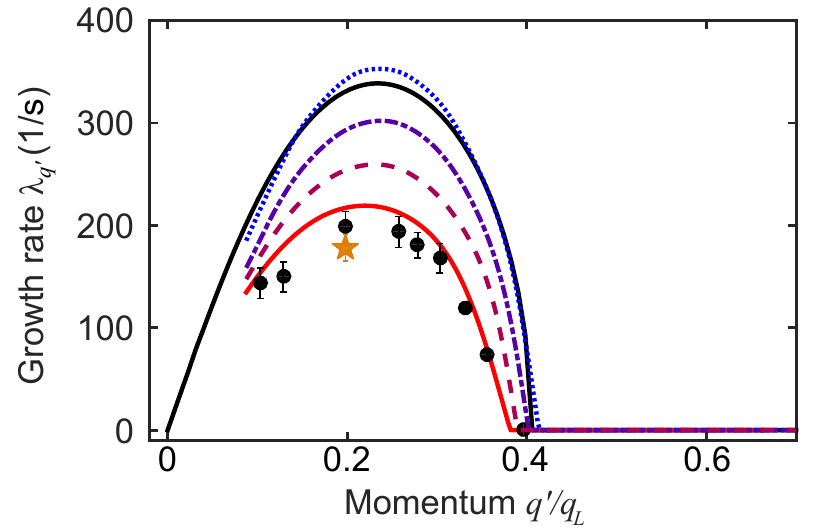}
\textbf{Fig.~S1} \textbf{Simulation of inflaton growth rate}. Here we show the measured growth rate $\lambda_{q'}$ from seeded (black dot) as well as the unseeded quench experiment (orange star). Black solid line is the prediction from Bogliubov theory with effective chemical potential $\mu$~=~$h\times$125 Hz. The lines show the results from simulation in color from blue to red with the seed amplitude $\delta\phi$: 0.1 (dot), 0.3 (dash-dot), 0.5 (dash) and 0.7 (solid) rad.
In the simulation, the trapping frequency $\omega$~=~$2\pi\times$~12 Hz, chemical potential $\mu$~=~$h\times$150 Hz, lattice depth is 8.9 $E_R$ and shaking amplitude $s$~=~25~nm, which are based on the corresponding experiment.
\end{figure}

Simulations with respect to different seed amplitudes explain the discrepancy between experiment and Bogliubov theory visible in Fig.~3~C from the main text. Here we show the inflaton growth rate $\lambda_{q'}$ from our simulation in Fig.~S1. For comparison, we also include the experimental data as well as the prediction from Bogliubov theory. First, for small initial population depletion from the $q$~=~0 state where $\delta\phi\leq$ 0.1, the reults from simulation agree well with that from Bogliubov theory. However, further increase of $\delta\phi$ quickly drive the system out of the perturbative regime. As a result, the growth rate $\lambda_{q'}$ decreases as the seed amplitude $\delta\phi$ increases. 
The measured growth rates concur with the simulation with seed amplitude $\delta\phi$ = 0.7, which is estimated to be 0.6(0.1) in experiments from the initial population depletion of the $q$~=~0 state.


\begin{thebibliography}{10}
\bibitem{Subir2011}
\bibinfo{author}{Sachdev, S.}
\newblock \emph{\bibinfo{title}{Quantum phase transitions}}
  (\bibinfo{publisher}{Cambridge University Press}, \bibinfo{year}{2011}),
  \bibinfo{edition}{1} edn.

\bibitem{Morikawa1995}
\bibinfo{author}{Morikawa, M.}
\newblock \bibinfo{title}{Cosmological inflation as a quantum phase
  transition}.
\newblock \emph{\bibinfo{journal}{Progress of Theoretical Physics}}
  \textbf{\bibinfo{volume}{93}}, \bibinfo{pages}{685--709}
  (\bibinfo{year}{1995}).

\bibitem{Kibble1980}
\bibinfo{author}{Kibble, T. W.~B.}
\newblock \bibinfo{title}{{Some Implications of a Cosmological Phase
  Transition}}.
\newblock \emph{\bibinfo{journal}{PHYSICS REPORTS (Review Section of Physics
  Letters)}} \textbf{\bibinfo{volume}{67}} (\bibinfo{year}{1980}).

\bibitem{Vojta2013}
\bibinfo{author}{Vojta, T.}
\newblock \bibinfo{title}{Quantum phase transitions}.
\newblock \emph{\bibinfo{journal}{AIP Conference Proceedings}}
  \textbf{\bibinfo{volume}{1550}}, \bibinfo{pages}{288--247}
  (\bibinfo{year}{2013}).

\bibitem{Guth2007}
\bibinfo{author}{Guth, A., M.}
\newblock \bibinfo{title}{Eternal inflation and its implications}.
\newblock \emph{\bibinfo{journal}{J. Phys. A: Math. Theor.}}
  \textbf{\bibinfo{volume}{40}}, \bibinfo{pages}{6811--6826}
  (\bibinfo{year}{2007}).

\bibitem{Kibble1976}
\bibinfo{author}{Kibble, T. W.~B.}
\newblock \bibinfo{title}{Topology of cosmic domains and strings}.
\newblock \emph{\bibinfo{journal}{J. Phys. A: Math. Gen.}}
  \textbf{\bibinfo{volume}{9}}, \bibinfo{pages}{1387--1398}
  (\bibinfo{year}{1976}).

\bibitem{Zurek1985}
\bibinfo{author}{Zurek, H., W.}
\newblock \bibinfo{title}{Cosmological experiments in superfluid helium?}
\newblock \emph{\bibinfo{journal}{Nature}} \textbf{\bibinfo{volume}{317}},
  \bibinfo{pages}{505--508} (\bibinfo{year}{1985}).

\bibitem{Zurek2014}
\bibinfo{author}{del Campo, A.} \& \bibinfo{author}{Zurek, W.~H.}
\newblock \bibinfo{title}{Universality of phase transition dynamics:
  Topological defects from symmetry breaking}.
\newblock \emph{\bibinfo{journal}{International Journal of Modern Physics A}}
  \textbf{\bibinfo{volume}{29}}, \bibinfo{pages}{1430018}
  (\bibinfo{year}{2014}).

\bibitem{Sadler2006}
\bibinfo{author}{Sadler, L.~E.}, \bibinfo{author}{Higbie, J.~M.},
  \bibinfo{author}{Leslie, S.~R.}, \bibinfo{author}{Vengalattore, M.} \&
  \bibinfo{author}{Stamper-Kurn, D.~M.}
\newblock \bibinfo{title}{Spontaneous symmetry breaking in a quenched
  ferromagnetic spinor bose-einstein condensate}.
\newblock \emph{\bibinfo{journal}{Nature}} \textbf{\bibinfo{volume}{443}},
  \bibinfo{pages}{312--315} (\bibinfo{year}{2006}).

\bibitem{Anatoli2011}
\bibinfo{author}{Polkovnikov, A.}, \bibinfo{author}{Sengupta, K.},
  \bibinfo{author}{Silva, A.} \& \bibinfo{author}{Vengalattore, M.}
\newblock \bibinfo{title}{Colloquium}.
\newblock \emph{\bibinfo{journal}{Rev. Mod. Phys.}}
  \textbf{\bibinfo{volume}{83}}, \bibinfo{pages}{863--883}
  (\bibinfo{year}{2011}).

\bibitem{Bloch2008}
\bibinfo{author}{Bloch, I.}, \bibinfo{author}{Dalibard, J.} \&
  \bibinfo{author}{Zwerger, W.}
\newblock \bibinfo{title}{Many-body physics with ultracold gases}.
\newblock \emph{\bibinfo{journal}{Rev. Mod. Phys.}}
  \textbf{\bibinfo{volume}{80}}, \bibinfo{pages}{885--964}
  (\bibinfo{year}{2008}).

\bibitem{Dziarmaga2010}
\bibinfo{author}{Dziarmaga, J.}
\newblock \bibinfo{title}{Dynamics of a quantum phase transition and relaxation
  to a steady state}.
\newblock \emph{\bibinfo{journal}{Adv. Phys.}} \textbf{\bibinfo{volume}{59}},
  \bibinfo{pages}{1063--1189} (\bibinfo{year}{2010}).

\bibitem{Baumann2011}
\bibinfo{author}{Baumann, K.}, \bibinfo{author}{Mottl, R.},
  \bibinfo{author}{Brennecke, F.} \& \bibinfo{author}{Esslinger, T.}
\newblock \bibinfo{title}{Exploring symmetry breaking at the dicke quantum
  phase transition}.
\newblock \emph{\bibinfo{journal}{Phys. Rev. Lett.}}
  \textbf{\bibinfo{volume}{107}}, \bibinfo{pages}{140402}
  (\bibinfo{year}{2011}).

\bibitem{Barnett2011}
\bibinfo{author}{Barnett, R.}, \bibinfo{author}{Polkovnikov, A.} \&
  \bibinfo{author}{Vengalattore, M.}
\newblock \bibinfo{title}{Prethermalization in quenched spinor condensates}.
\newblock \emph{\bibinfo{journal}{Phys. Rev. A}} \textbf{\bibinfo{volume}{84}},
  \bibinfo{pages}{023606} (\bibinfo{year}{2011}).

\bibitem{Lamporesi2013}
\bibinfo{author}{Lamporesi, G.}, \bibinfo{author}{Donadello, S.},
  \bibinfo{author}{Serafini, S.}, \bibinfo{author}{Dalfovo, F.} \&
  \bibinfo{author}{Ferrari, G.}
\newblock \bibinfo{title}{Spontaneous creation of kibble-zurek solitons in a
  bose-einstein condensate}.
\newblock \emph{\bibinfo{journal}{Nat. Phys.}} \textbf{\bibinfo{volume}{9}},
  \bibinfo{pages}{656--660} (\bibinfo{year}{2013}).

\bibitem{Nicklas2015}
\bibinfo{author}{Nicklas, E.} \emph{et~al.}
\newblock \bibinfo{title}{Observation of scaling in the dynamics of a strongly
  quenched quantum gas}.
\newblock \emph{\bibinfo{journal}{Phys. Rev. Lett.}}
  \textbf{\bibinfo{volume}{115}}, \bibinfo{pages}{245301}
  (\bibinfo{year}{2015}).

\bibitem{Navon2015}
\bibinfo{author}{Navon, N.}, \bibinfo{author}{Gaunt, A.~L.},
  \bibinfo{author}{Smith, R.~P.} \& \bibinfo{author}{Hadzibabic, Z.}
\newblock \bibinfo{title}{Critical dynamics of spontaneous symmetry breaking in
  a homogeneous bose gas}.
\newblock \emph{\bibinfo{journal}{Science}} \textbf{\bibinfo{volume}{347}},
  \bibinfo{pages}{167--170} (\bibinfo{year}{2015}).

\bibitem{Klinder2015}
\bibinfo{author}{Klinder, J.}, \bibinfo{author}{Keßler, H.},
  \bibinfo{author}{Wolke, M.}, \bibinfo{author}{Mathey, L.} \&
  \bibinfo{author}{Hemmerich, A.}
\newblock \bibinfo{title}{Dynamical phase transition in the open dicke model}.
\newblock \emph{\bibinfo{journal}{Proceedings of the National Academy of
  Sciences}} \textbf{\bibinfo{volume}{112}}, \bibinfo{pages}{3290--3295}
  (\bibinfo{year}{2015}).

\bibitem{Meldgin2016}
\bibinfo{author}{Meldgin, C.} \emph{et~al.}
\newblock \bibinfo{title}{Probing the bose glass-superfluid transition using
  quantum quenches of disorder}.
\newblock \emph{\bibinfo{journal}{Nat. Phys.}} \textbf{\bibinfo{volume}{12}},
  \bibinfo{pages}{646--649} (\bibinfo{year}{2016}).

\bibitem{Anquez2016}
\bibinfo{author}{Anquez, M.} \emph{et~al.}
\newblock \bibinfo{title}{Quantum kibble-zurek mechanism in a spin-1
  bose-einstein condensate}.
\newblock \emph{\bibinfo{journal}{Phys. Rev. Lett.}}
  \textbf{\bibinfo{volume}{116}}, \bibinfo{pages}{155301}
  (\bibinfo{year}{2016}).

\bibitem{Clark2016}
\bibinfo{author}{Clark, L.~W.}, \bibinfo{author}{Feng, L.} \&
  \bibinfo{author}{Chin, C.}
\newblock \bibinfo{title}{Universal space-time scaling symmetry in the dynamics
  of bosons across a quantum phase transition}.
\newblock \emph{\bibinfo{journal}{Science}} \textbf{\bibinfo{volume}{354}},
  \bibinfo{pages}{606--610} (\bibinfo{year}{2016}).

\bibitem{Harry2013}
\bibinfo{author}{Parker, C.~V.}, \bibinfo{author}{Ha, L.-C.} \&
  \bibinfo{author}{Chin, C.}
\newblock \bibinfo{title}{Direct observation of effective ferromagnetic domains
  of cold atoms in a shaken optical lattice}.
\newblock \emph{\bibinfo{journal}{Nat Phys}} \textbf{\bibinfo{volume}{9}},
  \bibinfo{pages}{769--774} (\bibinfo{year}{2013}).

\bibitem{Harry2015}
\bibinfo{author}{Ha, L.-C.}, \bibinfo{author}{Clark, L.~W.},
  \bibinfo{author}{Parker, C.~V.}, \bibinfo{author}{Anderson, B.~M.} \&
  \bibinfo{author}{Chin, C.}
\newblock \bibinfo{title}{Roton-maxon excitation spectrum of bose condensates
  in a shaken optical lattice}.
\newblock \emph{\bibinfo{journal}{Phys. Rev. Lett.}}
  \textbf{\bibinfo{volume}{114}}, \bibinfo{pages}{055301}
  (\bibinfo{year}{2015}).

\bibitem{Shvarchuck2002}
\bibinfo{author}{Shvarchuck, I.} \emph{et~al.}
\newblock \bibinfo{title}{Bose-einstein condensation into nonequilibrium states
  studied by condensate focusing}.
\newblock \emph{\bibinfo{journal}{Phys. Rev. Lett.}}
  \textbf{\bibinfo{volume}{89}}, \bibinfo{pages}{270404}
  (\bibinfo{year}{2002}).

\bibitem{Chenlung2011}
\bibinfo{author}{Hung, C.~L.} \emph{et~al.}
\newblock \bibinfo{title}{Extracting density-density correlations from in situ
  images of atomic quantum gases}.
\newblock \emph{\bibinfo{journal}{Phys. Rev. A}} \textbf{\bibinfo{volume}{59}},
  \bibinfo{pages}{4595--4607} (\bibinfo{year}{1999}).

\bibitem{Morsch2006}
\bibinfo{author}{Morsch, O.} \& \bibinfo{author}{Oberthaler, M.}
\newblock \bibinfo{title}{Dynamics of bose-einstein condensates in optical
  lattices}.
\newblock \emph{\bibinfo{journal}{Rev. Mod. Phys.}}
  \textbf{\bibinfo{volume}{78}}, \bibinfo{pages}{179--215}
  (\bibinfo{year}{2006}).

\bibitem{Anglin2003}
\bibinfo{author}{Anglin, J.~R.}
\newblock \bibinfo{title}{Second-quantized landau-zener theory for dynamical
  instabilities}.
\newblock \emph{\bibinfo{journal}{Phys. Rev. A}} \textbf{\bibinfo{volume}{67}},
  \bibinfo{pages}{051601} (\bibinfo{year}{2003}).
  
  

\bibitem{Pethick2008}
\bibinfo{author}{Pethick, J., C} \& \bibinfo{author}{Smith, H.}
\newblock \emph{\bibinfo{title}{Bose-Einstein Condensation in Dilute Gases}}
  (\bibinfo{publisher}{Cambridge University Press}, \bibinfo{year}{2008}),
  \bibinfo{edition}{2} edn.


\bibitem{Landau2008}
\bibinfo{author}{Landau, L.~D.} \& \bibinfo{author}{Lafshitz, E.~M.}
\newblock \emph{\bibinfo{title}{Statistical Physics, Part 2}}
  (\bibinfo{publisher}{Oxford}, \bibinfo{year}{2008}), \bibinfo{edition}{3}
  edn.
  
\end{thebibliography}
\end{document}